\documentclass[11pt,a4paper]{article}

\usepackage[title]{appendix}
\usepackage[affil-sl,auth-sc,noblocks]{authblk}
\usepackage[noblocks]{authblk}
\usepackage[top=2cm, bottom=2cm, left=2cm, right=2cm]{geometry}
\usepackage{amssymb}
\usepackage{amsbsy}
\usepackage{mathrsfs,amsmath}
\usepackage{mathtools}
\usepackage{graphicx}
\usepackage{tensor}
\usepackage{array}
\usepackage{graphbox}
\usepackage{multicol}
\usepackage{cite}
\usepackage{appendix}
\graphicspath{{}}
\usepackage{enumitem}
\usepackage{soul}

\allowdisplaybreaks

\usepackage{caption}
\usepackage{hyperref}
\hypersetup{
	colorlinks=true,
	linkcolor=blue,
	citecolor=red
}
\numberwithin{equation}{section}

\usepackage{soul,color}

\title{Disentangling SMEFT and UV contributions in $h\to Z\gamma$ and $h\to\gamma\gamma$ decays}
\author{Kostas Mantzaropoulos}
\affil{Department of Physics, University of Ioannina, GR 45 110, Greece}
\date{\today}

\begin{document}
\maketitle

\begin{abstract}
LHC searches have revealed that the Higgs boson decay to a photon pair is nearly consistent with the Standard Model (SM), whereas recently, there is evidence for the decay of the Higgs boson to a $Z$-boson and a photon. These decays are governed by the same set of Wilson-coefficients at the tree level in Standard Model Effective Field Theory (SMEFT).
In this study, we aim to explain this potential discrepancy between the decays $h\to \gamma \gamma$ and $h\to Z\gamma$. We conduct a model-independent analysis in SMEFT to determine the magnitude and features of the Wilson coefficients required to account for the observed distinction between the two signal strengths. Following this, we adopt a top-down approach, considering all single and two field extensions of the SM, including scalars and fermions, as candidates for new interactions. We perform the matching of these models to one loop using automated packages and compare the models' predictions regarding $h\to Z\gamma$.
\end{abstract}

\section{Introduction}

Since the discovery of the Higgs particle by the LHC \cite{ATLAS:2012yve,CMS:2012qbp} and the absence of any new smoking gun event our attention is increasingly turning to studying in detail the properties and the couplings of the Higgs particle. Its decay and production can be affected by particles that have not yet been discovered or hinted at and are thus elusive. If these new particles are heavy the Standard Model Effective Field Theory (SMEFT) (for reviews we refer the reader to Refs. \cite{Henning:2014wua,  Brivio:2017vri, Isidori:2023pyp}) provides a framework to study the effects of particles that are found above the electroweak (EW) scale, namely $v \sim ~245 ~\rm{GeV}$. The constituents of this framework are higher dimensional operators that respect the SM gauge group, {thus} we add to the SM Lagrangian the sum of terms with mass dimension greater than 4,
\begin{align}
    \mathcal{L} = \mathcal{L}_{\rm{SM}} + \frac{C^{5} \mathcal{O}^{5}}{\Lambda} +\frac{C^{6} \mathcal{O}^{6}}{\Lambda^2} + \ldots\;,
\end{align}
where as a superscript the mass dimension of each respective operator is denoted and $\Lambda$ serves as the scale of new (UV) physics. Each operator is accompanied by its respective coefficient known as Wilson coefficient (Wc) whose expressions encode the effect of UV physics. 

Deviations from the SM values of decay and production channels are encoded in the so called signal strength. These are calculated as the ratio,
\begin{align}
    \mu_{h \to X} = \frac{\Gamma(\text{SMEFT}, h\to X)}{\Gamma(\text{SM}, h\to X)} = 1 + \frac{\Gamma(\text{BSM}, h\to X)}{\Gamma(\text{SM}, h\to X)}\;,
\end{align}
where with $\Gamma(\text{BSM})$ we denote the decay rate that is affected only by physics beyond the SM (BSM). The signal strengths are split in the following manner $\mu_{h\to X} = 1 +\delta R_{h\to X}$, where $\delta R_{h\to X} = \Gamma(\text{BSM}, h\to X)/\Gamma(\text{SM},h\to X)$. Hence new physics that contributes to each respective channel would shift $\mu_{h\to X}$ to values different than one. The decays of the Higgs into two bosons have been measured with high precision and in agreement with the SM are $WW^\ast, ZZ^\ast, \gamma\gamma$ \cite{ATLAS:2018xbv, ATLAS:2019vrd, CMS:2018zzl, CMS:2020dvg, CMS:2018gwt, ATLAS:2013dos, ATLAS:2018hxb}. A detailed analysis of the properties, the decay and production as well as future directions for Higgs physics can be found in \cite{CMS:2022dwd, ATLAS:2022vkf}. Although most of the aforementioned decays are in agreement with the SM a recent analysis by the ATLAS and the CMS collaboration \cite{ATLAS:2023yqk} found evidence for the decay of $h \to Z\gamma$, which was an elusive decay up to now. Additionally, they reported a mild excess of around $\sim 2\sigma$ with respect to the SM value, the measured signal strength is,
\begin{align}
    \mu_{h\to Z\gamma} = 2.2 \pm 0.7\;.
\end{align}
Subtracting the SM value of the signal strength $\mu^{\text{SM}}_{h\to Z\gamma}=1$, this implies that contributions from new physics must account for $\mu_{h\to Z\gamma}^{\text{BSM}} = 1.2 \pm 0.7$, under the assumption that all uncertainties originate from the UV sector. This study aims to answer this question, which UV model could account for such an excess in the $h \to Z \gamma$ decay? The answer lies in the values of the Wcs that are found in the SMEFT. In the SMEFT however, there is another Higgs decay that receives contributions from almost the same set of operators, the $h\to\gamma\gamma$ decay, and the expressions governing these observables are similar as it is explored in Section \ref{sec: section2}. Thus it is important to investigate these two decays under the same scope and try to disentangle the contributions that give an excess in $h\to Z\gamma$. A few recent articles have tried to address this question varying the field content or extending the gauge group of the SM \cite{He:2024sma,Boto:2023bpg,Hong:2023mwr,Barducci:2023zml}. Another interesting article explores enhancements of $h \to Z\gamma$ through renormalization group effects of tree-level generated dimension 8 operators including also massive vector-boson fields \cite{Grojean:2024tcw}, which are not considered in this work. Contributions from the MSSM in the $h\to Z\gamma$ can be found in the ref. \cite{Israr:2024ubp}. Additionally, the decay $h\to Z\gamma$ is proposed as a probe to test the compositeness of the Higgs boson, under the EFT framework, in \cite{Cao:2018cms}.

The values of the Wcs that can affect the properties of the Higgs particle can be calculated by the process known as matching. In the matching procedure (top-down approach) Wcs are calculated through a complete UV model and their analytic expressions depend on the coupling of the UV model as well as the various scales (or scale). This procedure can be done either diagrammatically, or directly through the path integral. Each respective approach carries advantages as well as disadvantages. Functional matching since its inception \cite{Gaillard:1985uh, Cheyette:1987qz} has recently seen renowned interest with new techniques \cite{Zhang:2016pja,Cohen:2020fcu} and universal results making their appearance \cite{Drozd:2015rsp, Ellis:2016enq, Fuentes-Martin:2016uol,Ellis:2017jns,Summ:2018oko,Kramer:2019fwz,Summ:2020dda,Ellis:2020ivx,Angelescu:2020yzf}, even at two-loop order \cite{Fuentes-Martin:2023ljp}. The automation of these techniques has evolved to encompass a broad spectrum of methods, from SuperTrace calculations \cite{Cohen:2020qvb, Fuentes-Martin:2020udw} to the efficient computation of Wilson coefficients (at both tree and loop level) directly from the Lagrangian \cite{DasBakshi:2018vni, Fuentes-Martin:2022jrf, Carmona:2021xtq, Guedes:2023azv, Dedes:2021abc}.

Another approach to determining the effect of new physics on observables is the bottom-up approach. This constitutes a model-agnostic method where the Wcs remain unknown and their value is determined by fitting a set of observables to a set of coefficients. Throughout this article we will be using the Warsaw basis \cite{Grzadkowski:2010es} to study the effects of UV independent Wcs to the relevant decays of the Higgs boson. Operators only up to dimension-6 will be considered, whereas the only dimension-5 operator contained in the Warsaw basis relates to neutrino masses and is irrelevant to the observables that are considered here. This allows us to gauge the magnitude of new physics that may affect each Wc respectively and may give us hints about the structure of the UV model. However the downside of this approach is the unknown correlations between Wcs that may arise from the UV model, so the bottom-up and the top-down approach should be seen as complementary to each other. One may give hints on high-scale physics while the other clearly determines the correlations of Wcs and restricts the available space of Wcs since the couplings in the UV will be less than the independent Wcs in the Warsaw basis.

The structure of the paper is the following. In Section \ref{sec: section2} we study the Higgs decays $h \to \gamma\gamma$ and $h \to Z \gamma$ at one-loop order, in the bottom-up approach, in the context of SMEFT. We make several rescalings to Wcs in order to account for their tree and one-loop level counterparts. After clearing up major contributions to their signal strengths we fit all Wcs to a set of observables related to the Higgs sector and compare the values of Wcs needed to account for their observed values. In Section \ref{sec: section3}, our attention is shifted to possible UV complete models that may account for the values of Wcs found in the fit. All colorless single field extensions of the SM, that respect the gauge group, tabulated in the tree-level dictionary \cite{deBlas:2017xtg}, are considered. In Section \ref{sec: section4}, we categorize interactions with respect to their loop functions and devise a scheme to tabulate two-field models that generate Wcs relevant to the Higgs decays. We match all models to the Warsaw basis through automated packages and perform a constrained minimization to find the best values for the couplings and masses of each respective model. We leave the signal strength of the decay $h\to Z\gamma$ as a prediction of the model and plot the values for each one compared to $h \to\gamma\gamma$.

\section{Model independent analysis in SMEFT}
\label{sec: section2}

There are two Higgs decays in SMEFT that are of interest to us currently, namely, $h \to \gamma \gamma$ and $h \to Z \gamma$. Their semi-numerical expressions, at one-loop order, of the signal strengths, in the input scheme $\{G_{F},\, M_{W},\, M_{Z}\}$, and in units of $\text{TeV}^{-2}$, are \cite{Dedes:2018seb, Dedes:2019bew, Dawson:2018pyl},

\begin{align}
    &\delta R_{h \to Z \gamma} \simeq 0.18 \left(C^{\ell \ell}_{1221} -C^{\phi \ell (3)}_{11} -C^{\phi \ell (3)}_{22} \right) + 0.12 \left( C^{\phi\Box} -C^{\phi D} \right) \nonumber\\
    &- 0.01 \left( C^{d\phi}_{33} - C^{u\phi}_{33} \right) + 0.02 \left( C^{\phi u}_{33} +C^{\phi q(1)}_{33} -C^{\phi q(3)}_{33} \right)\nonumber\\
    &+\left[14.99 - 0.35 \log\frac{\mu^2}{M_{W}^{2}} \right] C^{\phi B} - \left[14.88 - 0.15 \log\frac{\mu^2}{M_{W}^{2}} \right] C^{\phi W} +\left[9.44 - 0.26 \log\frac{\mu^2}{M_{W}^{2}} \right] C^{\phi WB}\nonumber\\
    &+ \left[0.10 - 0.20 \log\frac{\mu^2}{M_{W}^{2}} \right] C^{W} -\left[0.11 - 0.04 \log\frac{\mu^2}{M_{W}^{2}} \right] C^{uB}_{33} + \left[0.71 - 0.28 \log\frac{\mu^2}{M_{W}^{2}} \right] C^{uW}_{33}\nonumber\\
    &-0.01\, C^{uW}_{22} - 0.01\,C^{dW}_{33} + \ldots\;,
    \label{eq:Zgam}
\end{align}

\begin{align}
     &\delta R_{h \to \gamma \gamma} \simeq 0.18 \left(C^{\ell \ell}_{1221} -C^{\phi \ell (3)}_{11} -C^{\phi \ell (3)}_{22} \right) + 0.12 \left( C^{\phi\Box} - 2C^{\phi D} \right) \nonumber\\
    &- 0.01 \left( C^{e\phi}_{22} + 4C^{e\phi}_{33}  +5C^{u\phi}_{22} +2C^{d\phi}_{33} -3C^{u\phi}_{33} \right)\nonumber\\
    &-\left[48.04 - 1.07 \log\frac{\mu^2}{M_{W}^{2}} \right] C^{\phi B} - \left[14.29 - 0.12 \log\frac{\mu^2}{M_{W}^{2}} \right] C^{\phi W} +\left[26.17 - 0.52 \log\frac{\mu^2}{M_{W}^{2}} \right] C^{\phi WB}\nonumber\\
    &+ \left[0.16 - 0.22 \log\frac{\mu^2}{M_{W}^{2}} \right] C^{W} +\left[2.11 - 0.84 \log\frac{\mu^2}{M_{W}^{2}} \right] C^{uB}_{33} + \left[1.13 - 0.45 \log\frac{\mu^2}{M_{W}^{2}} \right] C^{uW}_{33}\nonumber\\ &-\left[0.03 - 0.01 \log\frac{\mu^2}{M_{W}^{2}} \right] C^{uB}_{22}
    - \left[0.01 - 0.00 \log\frac{\mu^2}{M_{W}^{2}} \right] C^{uW}_{22}
    +\left[0.03 - 0.01 \log\frac{\mu^2}{M_{W}^{2}} \right] C^{dB}_{33}\nonumber\\
    &-\left[0.02 - 0.01 \log\frac{\mu^2}{M_{W}^{2}} \right] C^{dW}_{33} +\left[0.02 - 0.00 \log\frac{\mu^2}{M_{W}^{2}} \right] C^{eB}_{33} - \left[0.01 - 0.00 \log\frac{\mu^2}{M_{W}^{2}} \right] C^{eW}_{33} + \ldots\;,
    \label{eq:gamgam}
\end{align}

where the dots denote terms whose contributions are lower than $0.01\times C$ and $\mu$ is the renormalization scale. We are restricting ourselves to contributions without CP-violation. These operators are heavily constrained by Electric Dipole Moments (EDMs), and would contribute with tiny corrections to the observables and the problem we are trying to tackle in this work. 

At a first glance we can naively say that the main contributions in these two observables originate from the same three operators, $\left\{C^{\phi B}, C^{\phi W}, C^{\phi WB}\right\}$,
however in the tree-level dictionary \cite{deBlas:2017xtg} these operators arise only from one-loop processes, assuming that the UV-Lagrangian contains terms only up to dimension 4, and restrict ourselves to UV models containing only scalars and/or fermions. {If we consider the presence of operators with dimension greater than 4 in the UV Lagrangian, or if we incorporate vector fields into the analysis, these operators can also be generated at tree level.

One way to disentangle the tree and loop level contributions is to split the coefficients into their tree and loop level parts as follows, $C = C^{[0]} + \frac{1}{16\pi^2} C^{[1]} \simeq C^{[0]} + 0.6 \times 10^{-2} C^{[1]}$. From what was discussed in the previous paragraph, we shall set $C^{[0]\phi B}$ , $C^{[0]\phi W}$ and $C^{[0]\phi WB}$ to zero. This re-scales all coefficients so that we can easily compare between contributions of different operators. Apart from the tree-loop split, for the case of the three Wilson coefficients mentioned above, we can do another re-scaling, $C^{\phi B} \to g^{\prime2} \hat{C}^{\phi B}$, $C^{\phi W} \to g^{2} \hat{C}^{\phi W}$, $C^{\phi WB} \to g^{\prime}g \hat{C}^{\phi WB}$. We immediately see that in both expressions the operators $C^{(e,u,d)(B,W)}$, $C^{W}$, are small since they occur at one-loop in the SMEFT. We can also observe that the combination of the three Wcs $C^{\ell\ell}_{1221}$, $C^{\phi\ell(3)}_{11,22}$ constitutes the correction of SMEFT to the Fermi constant, which is known to high accuracy. The corrections of the Fermi constant in the SMEFT read \cite{Descotes-Genon:2018foz,Dedes:2017zog}, $G_{F}^{\rm{SMEFT}} = G_{F} + \delta G_{F}$, where $G_{F} = 1.1663787(6)\times 10^{-5}\,\rm{GeV}^{-2}$ and $\delta G_{F} = -\frac{1}{\sqrt{2}}\,\left(C^{\ell \ell}_{1221} -C^{\phi \ell (3)}_{11} -C^{\phi \ell (3)}_{22} \right)$.
So, we re-write the formulas with the re-scaled contributions, setting the renormalization scale to $\mu = \Lambda = 1~\text{TeV}$ and the Fermi constant correction, in order to compare the coefficients again,

\begin{align}
    \delta R_{h \to Z \gamma} &\simeq -0.25\, \delta G_{F} + 0.12 \left( C^{[0]\phi\Box} -C^{[0]\phi D} \right)\nonumber\\
    &+0.01 \hat{C}^{[1]\phi B} -0.04\, \hat{C}^{[1]\phi W} + 0.01\,\hat{C}^{[1]\phi WB} \nonumber\\
    &- 0.01 \left( C^{[0]d\phi}_{33} - C^{[0]u\phi}_{33} \right) + \ldots \;,\\
    \nonumber\\
    \delta R_{h \to \gamma \gamma} &\simeq -0.25\, \delta G_{F} + 0.12 \left( C^{[0]\phi\Box} -2 C^{[0]\phi D} \right) \nonumber\\
    &-0.03\, \hat{C}^{[1]\phi B} -0.04\, \hat{C}^{[1]\phi W} +0.03\, \hat{C}^{[1]\phi WB} \nonumber\\
    &- 0.01 \left( C^{[0]e\phi}_{22} + 4C^{[0]e\phi}_{33}\right)\nonumber\\
    &-0.01 \left(5C^{[0]u\phi}_{22} +2C^{[0]d\phi}_{33} -3C^{[0]u\phi}_{33} \right) + \ldots\;.
\end{align}
Disentangling the formulas in such a way provides us with more insight on the values of the couplings that we can expect in the UV. With this procedure we have also narrowed down a multitude of contributing operators to just a handful of them making the study of these two observables easier.

A few remarks are in order.
\begin{itemize}
    \item The largest contribution to $\delta R_{h \to \gamma \gamma}$ no longer originates from the operator $C^{\phi B}$ even though initially that was the case. The most dominant Wcs are $C^{\ell \ell}_{1221}$, $C^{\phi \ell (3)}_{11}$, $C^{\phi \ell (3)}_{22}$, while next in magnitude are the Wcs $C^{\phi D}$ and $C^{\phi\Box}$. Incidentally, these five operators contribute maximally to $\delta R_{h \to Z \gamma}$ as well.
    \item The model in question must not generate large corrections to the Fermi constant, whose SMEFT expression depends on $C^{\ell \ell}_{1221}$, $C^{\phi \ell (3)}_{11}$, $C^{\phi \ell (3)}_{22}$, which is excluded, unless we tune down its couplings to minuscule values to reach a correction of the order of $10^{-6}$. Additionally, Wilson coefficients $C^{\phi\Box}$ and $C^{\phi D}$ could cancel each other out in $h\to\gamma\gamma$ if $C^{\phi\Box} = 2 C^{\phi D}$ and that could boost $h\to Z\gamma$, unless they are not generated at tree level and are suppressed by a loop factor. However, in the case of a cancellation both $C^{\phi D}$ and $C^{\phi \Box}$ would be heavily constrained by the $T$-parameter, since the coefficients are directly related to each other.
    \item What values of Wcs would it take to boost $h \to Z \gamma$ while simultaneously these Wcs would destructively contribute in $h \to \gamma \gamma$? Ideally we would like to avoid generating the dominant tree level Wcs mentioned in the two previous bullets because they equally contribute to both observables and there is no apparent way to cancel each other out. Our main goal is to restore the dominance of $C^{\phi(B,W,WB)}$ Wcs.
    \item For both signal strengths their expressions are almost identical with the only difference being the Wc $C^{e\phi}_{pp}$ which contributes only to $\delta R_{h \to \gamma \gamma}$, however operators of such kind arising from the tree level are usually suppressed by the Yukawa coupling of the corresponding fermion and are thus suppressed for the most part. For this reason we will keep from now on only contributions from the third generation of quarks.
    \item In the expressions in eq.(\ref{eq:Zgam}) and eq.(\ref{eq:gamgam}) for the signal strengths all Wcs are considered at the renormalization scale $\mu$, $C(\mu)$. Since we have split the operators in the tree and loop counterparts the RG mixing of loop level operators constitutes a two loop effect and is neglected. Since we have set $\mu = \Lambda = 1~\text{TeV}$, all Wcs are from now on considered at $C(\mu) = C(\Lambda)$.
\end{itemize}

We construct a chi-square function to explore further the correlations and the required numerical values that these three coefficients need to take to accommodate an excess in one over the other observable. The observables that we choose to add are the decay and production signal strengths of the Higgs boson which can fairly constrain all Wcs in our study. We consider the decays of the Higgs boson tabulated in the first column of Table \ref{tab:numerics} and the for the production modes we include gluon fusion (ggF), vector boson fusion (VBF), associated production with a vector boson (Wh, Zh) and lastly, associated production with a pair of top quakrs (tth). Apart from these we also add the oblique parameters $S$ and $T$, since they highly constrain $C^{\phi WB}$ and $C^{\phi D}$ respectively. In the $\{G_{F}, M_{W}, M_{Z}\}$ scheme, these two expressions read,
\begin{align}
    \Delta S &= \frac{2 \pi}{G_{F}^{2} M_{W} \sqrt{M_{Z}^2 - M_{W}^2}}\, \frac{C^{\phi WB}}{\Lambda^2}\;,\\
    \Delta T &= \frac{\pi}{4 G_F^2}\,\frac{M_Z^2}{M_W^2 (M_Z^2 -M_W^2)}\,\frac{C^{\phi D}}{\Lambda^2}
\end{align}
Substituting the relevant values and setting $\Lambda = 1\, \rm{TeV}$, we get the semi-numerical expression,
\begin{align}
    \Delta S = 0.0199\, \hat{C}^{[1]\phi WB}\;,\\
    \Delta T = -4.0083\, C^{\phi D}\;.
\end{align}
The experimental values of decay and production channels that we are using to construct the $\chi^2$ function are shown in Table \ref{tab:numerics}. For $S$ and $T$ parameters we have the following two experimental values and corresponding uncertainties, $S_{\rm{exp}} = -0.02\pm 0.07$ and $T_{\rm{exp}} = 0.04\pm 0.06$ \cite{Workman:2022ynf}.

\begin{table}[ht]
    \centering
    \begin{tabular}{|cc|cc|}
        \hline
        Decay & Experiment & Production & Experiment \\
        \hline
        $\delta R_{h \to \gamma\gamma}$ & $0.10\pm 0.07$ & $\delta R_{ggF}$ & $-0.03\pm 0.08$\\
        $\delta R_{h \to Z\gamma}$ & $1.20\pm 0.70$ \cite{ATLAS:2023yqk} & $\delta R_{VBF}$ & $-0.20\pm 0.12$\\
        $\delta R_{h \to ZZ^\ast}$ & $0.02\pm 0.08$ & $\delta R_{Wh}$ & $0.44\pm 0.26$\\
        $\delta R_{h \to WW^\ast}$ & $0.00\pm 0.08$ & $\delta R_{Zh}$ & $0.29\pm 0.25$\\
        $\delta R_{h \to \mu^+ \mu^-}$ & $0.21\pm 0.35$ & $\delta R_{tth}$ & $-0.06\pm 0.20$\\
        $\delta R_{h \to \tau^+ \tau^-}$ & $-0.09\pm 0.09$ & &\\ 
        $\delta R_{h \to b\bar{b}}$ & $0.01\pm 0.12$ & &\\
        \hline
    \end{tabular}
    \caption{Numerical values used to construct the $\chi^2$ function to be minimized. The decay channel values were taken from \cite{Workman:2022ynf}, except for the $h \to Z \gamma$ decay. The production channel experimental values were taken from CMS\cite{CMS:2022dwd}.}
    \label{tab:numerics}
\end{table}
We define the $\chi^2$,
\begin{align}
    \chi^2 = \left( O^{\rm{SMEFT}} - O^{\rm{exp}} \right)^{T} \left(\sigma^2\right)^{-1} \left( O^{\rm{SMEFT}} - O^{\rm{exp}} \right)\;,
\end{align}
where $O^{\rm{exp}}$ is the column vector that contains the central values of the experimental measurements of the corresponding signal strengths of Table \ref{tab:numerics}, while in this case $\sigma^2$ is a $N_{\rm{obs}}\times N_{\rm{obs}}$ diagonal matrix containing the relevant uncertainties, where we have neglected theory uncertainties and assumed that all measurements are uncorrelated. The quantity $O^{\rm{SMEFT}}$ can be decomposed into two pieces, a purely SM piece, $O^{\rm{SM}}$ and a purely BSM piece, $O^{\rm{BSM}}$. Since the SM piece is a pure number we subtract it from the experimental values and we define $O^{\delta} = O^{\rm{SM}} - O^{\rm{exp}}$. We use Singular Values Decomposition (SVD) technique as described in \cite{Bodwin:2019ivc} to solve this least squares problem and we cross-check the results by also minimizing the chi-square function. The set of Wcs we consider here are
\begin{align}
    \left\{C^{\phi B},~ C^{\phi W},~ C^{\phi WB},~ C^{\phi\Box},~ C^{\phi D},~ C^{u\phi}_{33},~ C^{d\phi}_{33}\right\}\;,
\end{align}
where we neglect the Wcs that affect corrections to the Fermi constant.

The best-fit values for the Wilson coefficients that we obtain, for $\Lambda = 1\,\rm{TeV}$ are,
\begin{align}
    \hat{C}^{[1]\phi B} &= 23.75 \pm 3.03\;,\\
    \hat{C}^{[1]\phi W} &= -26.26 \pm 2.68\;,\\
    \hat{C}^{[1]\phi WB} &= -1.07 \pm 3.63\;,\\
    C^{\phi \Box} &= -0.18 \pm 0.40\;,\\
    C^{\phi D} &= 0.01 \pm 0.01\;,\\
    C^{u\phi}_{33} &= -0.83 \pm 1.21\;,\\
    C^{d\phi}_{33} &= 0.001 \pm 0.03\;.
\end{align}
We also note the values of the first three Wcs before rescalings to be, $C^{\phi B} = 0.018\pm 0.002$, $C^{\phi W} = -0.071 \pm 0.007$ and $C^{\phi WB} = -0.0015 \pm 0.005$. Comparing the results of our SMEFT analysis it can be seen that within the marginalised values reported in refs. \cite{Ellis:2020unq, Ethier:2021bye} our own values fall within the ranges given in the aforementioned refs. However, if we consider the individual cases where only one Wc contributes to observables the values that we obtain fall short of the ranges obtained in the global fits. This deviation could be justified by the amount of observables contained in the global fits, while in this study we focus dominantly on the Higgs sector observables.

The correlation matrix for these coefficients, is,
\begin{align}
    \begin{array}{c|ccccccc}
    \rm{Wcs} & C^{\phi B} & C^{\phi W} & C^{\phi WB} & C^{\phi \Box} & C^{\phi D} & C^{u\phi}_{33} & C^{d\phi}_{33}\\
     \hline
     C^{\phi B} & 1. & 0.268 & 0.633 & 0.073 & -0.013 & 0.192 & 0.032 \\
     C^{\phi W} &  & 1. & 0.613 & 0.539 & -0.007 & 0.335 & 0.205 \\
     C^{\phi WB} &  &  & 1. & 0.003 & 0 & 0.001 & 0.001 \\
     C^{\phi \Box} & & & & 1. & 0.03 & 0.33 & 0.374 \\
     C^{\phi D} & & & & & 1. & 0.006 & 0.007 \\
     C^{u\phi}_{33} & & & & & & 1. & 0.157 \\
     C^{d\phi}_{33} & & & & & & & 1. \\
    \end{array}
\end{align}

From the covariance matrix of the coefficients we can also draw the error ellipses, as shown in Figure \ref{fig:3_coef_ellipses}. In each of the plots we have set each respective coefficient that is not drawn to its best-fit values and have drawn error ellipses of the other two. For example, in the first plot we have set every coefficient but $C_{\phi W}$ and $C_{\phi WB}$ to is best-fit value and have plotted the error ellipses of the other two. 

\begin{figure*}
    \includegraphics[scale=0.45]{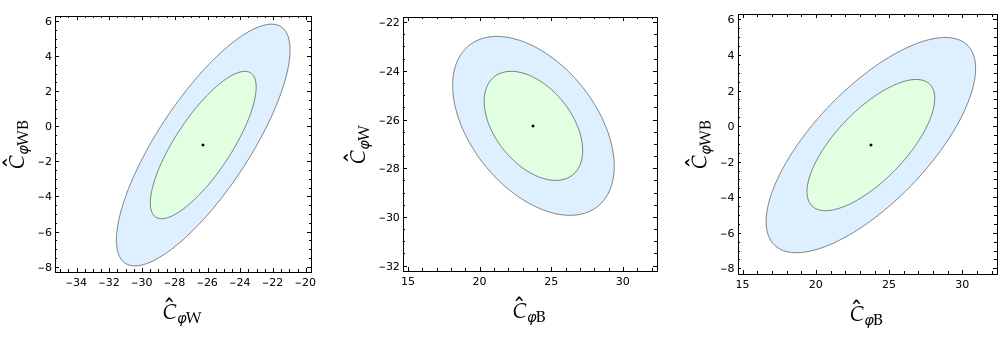}
    \caption{Error ellipses drawn by setting each respective coefficient to its best-fit value while varying the other two. Green and light blue contours represent, 1$\sigma$ and $2\sigma$ respectively.}
    \label{fig:3_coef_ellipses}
\end{figure*}

Let us now establish the criteria that UV models must meet in order to be considered viable for further study:
\begin{itemize}
    \item[1.] The most important relation that needs to hold is that $C^{\phi B}$ and $C^{\phi W}$ need to have a definite sign difference. This means that the UV model has to generate both. It alone could accommodate for an excess in $h\to Z \gamma$ if no other Wc that affects the two observables equally is generated.
    \item[2.] We also notice that if $C^{\phi WB}$ is generated it's value would need to be small, even accounting for uncertainties because of the direct relation to the $S$-parameter.
    \item[3.] Some models could also generate a pair of $C^{\phi\Box}$ and $C^{\phi D}$, if they are generated together at tree level they are proportional to each other. From the fit we see that including uncertainties it could be possible to accommodate this possibility. 
\end{itemize}

\section{Single field UV models}
\label{sec: section3}

In the following sections we examine the single field extensions of the tree level dictionary \cite{deBlas:2017xtg, Li:2023cwy} as well as some potential two field cases. To get all the expressions of the relevant generated operators we use the package SOLD \cite{Guedes:2023azv}, which contains all information on Wcs exclusively generated at one-loop order from scalar and vector like fermion extensions, while for tree-level generated operators we use the package MatchMakerEFT \cite{Carmona:2021xtq} and cross-check with the expressions found in ref. \cite{deBlas:2017xtg}. We refrain from including or discussing massive vector bosons for several reasons. First, no automated package exists up to date that could give us immediately Wcs and facilitate our study. Second and most important, we would need to account for the gauge boson mass, which would require the inclusion of a spontaneous symmetry breaking mechanism able to explain the origin of the mass. Additionally, if we consider vector bosons the Wcs that are of interest, namely $C^{\phi (B,W,WB)}$, are generated at tree level and are either coming from already non renormalizable interactions of dimension greater than four, or in some cases even if these operators are coming from interactions of dimension less or equal to four there is never a sign difference between Wcs $C^{\phi B}$ and $C^{\phi W}$ \cite{deBlas:2017xtg}. The inclusion of vector boson extensions is outside the scope of this study. Additionally, barring vector boson extensions, CP-violating operators come from interactions of already non-renormalizable operators in the models that we investigate, we restrict ourselves to interactions of the original Lagrangian up to mass dimension four.

\subsection{The case for BSM scalars}

To examine some viable scenarios we list in Table \ref{tab:scalar_ext} all scalar fields that serve as extensions of the SM, that also have a linear coupling with the SM particle content. We also list the tree level operators that they generate as well as the subset of loop level operators that are interesting for our case study. We exclude colored fields, as they induce the operator $\mathcal{O}^{\phi G}$, which significantly impacts the production rate of $ggF$. Given that this production channel is precisely measured, we aim to avoid constraints associated with this bound. In the following bullets all cases of Table \ref{tab:scalar_ext} will be investigated if they could account for the values of the Wcs obtained in the SMEFT analysis. In the matching calculations only the necessary operators for our discussion are provided, all other operators not presented are either zero or generated at loop-level and their contributions is deemed too small as has been discussed in Section \ref{sec: section2}. In the results presented below the notation of the tree-level dictionary \cite{deBlas:2017xtg} is being followed, the Lagrangian used for scalars can be found in Appendix A.2 of the aforementioned ref. For clarity however, the subset of the Lagrangian where the operator is being induced from is also presented. The mass terms of the new scalars are denoted by $M_{i}$, where $i = \left\{\mathcal{S},\,\mathcal{S}_1,\,\mathcal{S}_2,\,\varphi,\,\Xi,\,\Xi_1,\,\Theta_1,\,\Theta_3\right\}$.

\begin{table}[ht]
    \centering
    \begin{tabular}{cccc}
        Fields & Irrep & Tree level operators & Loop level operators \\
        \hline\\
         $\mathcal{S}$ & $(1,1)_{0}$ & $\mathcal{O}^{\phi\Box}$ & $\mathcal{O}_{\phi B}$, $\mathcal{O}_{\phi W}$, $\mathcal{O}_{\phi WB}$\\
         $\mathcal{S}_{1}$ & $(1,1)_{1}$ & $\mathcal{O}^{\ell\ell}$ & $\mathcal{O}_{\phi B}$ \\
         $\mathcal{S}_{2}$ & $(1,1)_{2}$ & $\mathcal{O}^{ee}$ & $\mathcal{O}_{\phi B}$\\
         $\varphi$ & $(1,2)_{1/2}$ & $\mathcal{O}^{(e,u,d)\phi}$ & $\mathcal{O}_{\phi B}$, $\mathcal{O}_{\phi W}$, $\mathcal{O}_{\phi WB}$\\
         $\Xi$ & $(1,3)_{0}$ & $\mathcal{O}^{\phi D}$, $\mathcal{O}^{\phi\Box}$, $\mathcal{O}^{(e,u,d)\phi}$ & $\mathcal{O}_{\phi B}$, $\mathcal{O}_{\phi W}$ \\
         $\Xi_{1}$ & $(1,3)_{1}$ & $\mathcal{O}^{\phi D}$, $\mathcal{O}^{\phi\Box}$, $\mathcal{O}^{(e,u,d)\phi}$ & $\mathcal{O}_{\phi B}$, $\mathcal{O}_{\phi W}$, $\mathcal{O}_{\phi WB}$\\
         $\Theta_{1}$ & $(1,4)_{1/2}$ & $\mathcal{O}^{\phi}$ & $\mathcal{O}_{\phi B}$, $\mathcal{O}_{\phi W}$, $\mathcal{O}_{\phi WB}$\\
         $\Theta_{3}$ & $(1,4)_{3/2}$ & $\mathcal{O}^{\phi}$ & $\mathcal{O}_{\phi B}$, $\mathcal{O}_{\phi W}$, $\mathcal{O}_{\phi WB}$\\
    \end{tabular}
    \caption{Tree and relevant loop level operators generated by new scalar field extensions of the SM. The first column follows the naming convention of ref. \cite{deBlas:2017xtg}, while in the second one the representation of each field is denoted as $\left(SU(3),\,SU(2)\right)_{U(1)}$.}
    \label{tab:scalar_ext}
\end{table}

\begin{itemize}
    \item Field $\mathcal{S}$, a neutral scalar generates all the necessary operators. The subset Lagrangian reads,
    \begin{align}
        \Delta\mathcal{L} \supset (\kappa_{\mathcal{S}})\,\mathcal{S}\,\phi^\dagger\phi 
    \end{align}
    The expressions of the Wcs are,
    \begin{align}
        &C^{\phi\Box} = -\frac{(\kappa_{\mathcal{S}})^2}{2 M_{\mathcal{S}}^4}\;,\\
        &\hat{C}^{[1]\phi W B} = 2\,\hat{C}^{[1]\phi B} = 2\,\hat{C}^{[1]\phi W} = \frac{(\kappa_{\mathcal{S}})^2}{6 M_{\mathcal{S}}^{4}}\;.
    \end{align}
    This set of Wcs would be impossible to explain a possible deviation because of the universal contribution to $C^{\phi\Box}$ and additionally $C^{\phi B} = C^{\phi W}$.
    \item Fields $\mathcal{S}_{1,2}$, charged singlets do no generate the required set of operators necessary for our purposes. Additionally, $\mathcal{S}_{1}$, gives $C_{\ell\ell}$ which would potentially contribute strongly to Fermi constant. For completeness, the expressions of the Wcs are
    \begin{align}
        \mathcal{S}_{1}:\quad &\left(C^{\ell\ell}\right)_{ijkl} = \frac{(y_{\mathcal{S}_1})^\ast_{jl}(y_{\mathcal{S}_1})_{ik}}{M_{\mathcal{S}_1}^2}\;,\\ &\hat{C}^{[1]\phi B} = -\frac{\lambda_{\mathcal{S}_{1}}}{12 M_{\mathcal{S}_1}^2}\;,\\
        \mathcal{S}_{2}:\quad &\left(C^{ee}\right)_{ijkl} = \frac{(y_{\mathcal{S}_2})^\ast_{lj}(y_{\mathcal{S}_2})_{ki}}{2M_{\mathcal{S}_2}^2}\;,\\ &\hat{C}^{[1]\phi B} = -\frac{\lambda_{\mathcal{S}_{2}}}{3 M_{\mathcal{S}_2}^2}\;,
    \end{align}
    where the Lagrangian for the relevant couplings reads,
    \begin{align}
        \Delta\mathcal{L} \supset \lambda_{\mathcal{S}_1}(\mathcal{S}_1^\dagger\mathcal{S}_1)(\phi^\dagger\phi) +\lambda_{\mathcal{S}_2}(\mathcal{S}_2^\dagger\mathcal{S}_2)(\phi^\dagger\phi)\nonumber\\
        (y_{\mathcal{S}_1})_{ij}\mathcal{S}_1^\dagger \bar{l}_{Li}i\sigma^2 l_{Lj}^c + (y_{\mathcal{S}_2})_{ij}\mathcal{S}_2^\dagger \bar{e}_{Ri} e_{Rj}^c + \text{h.c.}\;,
    \end{align}
    where $i,j$ are flavor indices.
    So, we exclude these two single fields as well.
    \item Next is the 2HDM, where a recent work \cite{DasBakshi:2024krs} also explores the matching of this model and fits to relevant Higgs observables. Field $\varphi$ generates the correct set. The interaction Lagrangian is presented below along with terms not contained in the tree-level dictionary (as were generated by SOLD),
    \begin{align}
        \Delta\mathcal{L} \supset &\kappa_{\varphi^2\phi^2}\, \phi^\dagger_\alpha \varphi^\dagger_\beta \phi_\gamma \varphi_\delta C^{(1)}_{\alpha\beta\gamma\delta} + \lambda_{\varphi^2\phi^2}\, \phi^\dagger_\alpha \varphi^\dagger_\beta \phi_\gamma \varphi_\delta C^{(2)}_{\alpha\beta\gamma\delta}\nonumber\\
        &(y_{e\varphi})_{ij}\varphi^\dagger\bar{e}_{Ri}l_{Lj} +(y_{d\varphi})_{ij}\varphi^\dagger\bar{d}_{Ri}q_{Lj} +\text{h.c.}\\
        &(y_{u\varphi})_{ij}\varphi^\dagger i\sigma^2 \bar{q}^T_{Li}u_{Rj} +\lambda_{\varphi}(\varphi^\dagger\phi)(\phi^\dagger\phi) + \text{h.c.}\;,
    \end{align}
    where a summation of the indices $\alpha,\beta,\gamma,\delta$ is implied, which represent SU(2) indices ranging $\alpha,\beta,\gamma,\delta = 1,2$, while $C$'s are the Clebsh-Gordan (CG) tensors of the coupling. The superscript T denotes transposition in SU(2) space. The non-zero elements of the corresponding CG tensors are,
    \begin{align}
        &C^{(1)}_{1111} = C^{(1)}_{2222} = 2\;,\\
        &C^{(1)}_{1212} = C^{(1)}_{1221} = C^{(1)}_{2121} = C^{(1)}_{2112} = 1\;,\\
        &C^{(2)}_{1212} = C^{(2)}_{2121} = -C^{(2)}_{1221} = -C^{(2)}_{2112} = 1\;.   
    \end{align}
    The expressions of the Wcs are,
    \begin{align}
        &\hat{C}^{[1]\phi B} = \hat{C}^{[1]\phi W} = -\frac{3\kappa_{\varphi^2\phi^2} + 2\lambda_{\varphi^2\phi^2}}{96\,M_{\varphi}^2}\;,\\
        &\hat{C}^{[1]\phi WB} = -\frac{\kappa_{\varphi^2\phi^2} - 2 \lambda_{\varphi^2\phi^2}}{48\,M_{\varphi}^2}\;,\\
        &\left(C^{e\phi}\right)_{ij} = \frac{\lambda_{\varphi}\,(y_{e\varphi})^\ast_{ji}}{M_{\varphi}^2}\;\\ &\left(C^{d\phi}\right)_{ij} = \frac{\lambda_{\varphi}\,(y_{d\varphi})^\ast_{ji}}{M_{\varphi}^2}\;\\ &\left(C^{u\phi}\right)_{ij} = -\frac{\lambda_{\varphi}^\ast\,(y_{u\varphi})_{ji}}{M_{\varphi}^2}\;. 
    \end{align}
    The Higgs-gauge boson operators are generated but the two most important ones are equal. For this reason we exclude this model.
    \item Next we consider electroweak triplets. First $\Xi$ generates the following operators,
    \begin{align}
        &C^{\phi\Box} = -\frac{1}{4}C^{\phi D} = \frac{\kappa^2_{\Xi}}{2M_{\Xi}^4}\;,\\
        &\hat{C}^{[1]\phi B} = \frac{\kappa^2_{\Xi}}{16 M_\Xi^4}\;,\\ &\hat{C}^{[1]\phi W} = -\frac{1}{4}\,\hat{C}^{[1]\phi B} + \frac{\lambda_{\Xi}}{6\sqrt{6} M_\Xi^2}\;,\\
        &\left(C^{e\phi}\right)_{ij} = \frac{\kappa^2_{\Xi}\,(y_{e})^\ast_{ji}}{M_{\Xi}^4}\;\\
        &\left(C^{d\phi}\right)_{ij} = \frac{\kappa^2_{\Xi}\,(y_{d})^\ast_{ji}}{M_{\Xi}^4}\;\\
        &\left(C^{u\phi}\right)_{ij} = -\frac{\kappa^2_{\Xi}\,(y_{u})^\ast_{ji}}{M_{\Xi}^4}\;,
    \end{align}
    where $y_{(e,u,d)}$ are the Yukawa coupling of the SM defined in eq.(A.1) of ref \cite{deBlas:2017xtg}. The corresponding Lagrangian for the rest of the couplings reads,
    \begin{align}
        \Delta\mathcal{L} \supset (\kappa_{\Xi})\,\phi^\dagger\Xi^{a}\sigma^{a}\phi + \lambda_{\Xi}\,\Xi^a\Xi^a\,(\phi^\dagger\phi)\;,
    \end{align}
    where a summation of the index $a$ is implied. The index denotes SU(2) triplets taking values $a=1,2,3$. Although the model appears promising as it generates all the necessary Wilson coefficients (Wcs), these Wcs are strongly correlated. We can express these correlations through the following relations:, $C^{\phi \Box} = -C^{\phi D}/4 = 8\,C^{\phi B}$, $(C^{(e,d)\phi})_{ij} = 2 (y_{(e,d)})^\ast_{ji}\,C^{\phi\Box}$ and $(C^{u\phi})_{ij} = -2 (y_u)^\ast_{ji}\,C^{\phi\Box}$. The most stringent constraint for this model comes from the $T$-parameter which is proportional to $C^{\phi D}$ and forces us to a small value for this Wc which is related to $C^{\phi B}$ which needs to have a large value. To provide a clearer picture, we can express the following observables, where we retain only the top Yukawa coupling and neglect other contributions due to their insignificance:
    \begin{align}
        \Delta T &= -8.016\frac{\kappa^2_{\Xi}}{M_\Xi^4}\;,\\
        \delta R_{h\to\gamma\gamma} &= 0.508\,\frac{\kappa^2_{\Xi}}{M_\Xi^4} - 0.0025\,\frac{\lambda_{\Xi}}{M_\Xi^2}\;,\label{eq: Xigg}\\
        \delta R_{h\to Z\gamma} &= 0.2912\,\frac{\kappa^2_{\Xi}}{M_\Xi^4} - 0.0026\,\frac{\lambda_{\Xi}}{M_\Xi^2}\;.\label{eq: XiZg}
    \end{align}
    In order to satisfy the bound for the $T$-parameter, which in our case needs to be negative and by taking the lower bound of the experimental value at $T_{\rm{exp}} \geq -0.02$, we get the bound for the coupling to mass ratio $ \kappa^2_{\Xi}/M_\Xi^4 \geq 25\times 10^{-4}$, substituting these values into eqs. (\ref{eq: Xigg}) and (\ref{eq: XiZg}) we get,
    \begin{align}
        \delta R_{h\to\gamma\gamma} &= 0.0013 - 0.0025\,\frac{\lambda_{\Xi}}{M_\Xi^2}\;,\\
        \delta R_{h\to Z\gamma} &= 0.0007 - 0.0026\,\frac{\lambda_{\Xi}}{M_\Xi^2}\;.
    \end{align}
    These contributions are insufficient, leaving only the universal effect of the triplet's quartic coupling with the Higgs, which cannot serve the purpose of splitting apart the two observables. We rule this model out as well.
    \item The next triplet, $\Xi_1$ is charged and the corresponding Wcs are,
    \begin{align}
        &C^{\phi\Box} = \frac{1}{2}C^{\phi D} = \frac{2|\kappa_{\Xi_1}|^2}{M_{\Xi_1}^4}\;,\\
        &\hat{C}^{[1]\phi WB} = -\frac{5|\kappa_{\Xi_1}|^2}{12 M_{\Xi_1}^4} -\frac{\lambda^\prime_{\Xi_1}}{6\sqrt{3} M_{\Xi_1}^2}\;,\\
        &\hat{C}^{[1]\phi B} = -\frac{|\kappa_{\Xi_1}|^2}{4 M_{\Xi_1}^4} +\frac{\lambda_{\Xi_1}}{4\sqrt{6} M_{\Xi_1}^2}\;,\\
        &\hat{C}^{[1]\phi W} = -\frac{|\kappa_{\Xi_1}|^2}{12 M_{\Xi_1}^4} +\frac{\lambda_{\Xi_1}}{6\sqrt{6} M_{\Xi_1}^2}\;,\\
        &\left(C^{e\phi}\right)_{ij} = \frac{2|\kappa_{\Xi_1}|^2\,(y_{e})^\ast_{ji}}{M_{\Xi_1}^4}\;,\\
        &\left(C^{d\phi}\right)_{ij} = \frac{2|\kappa_{\Xi_1}|^2\,(y_{d})^\ast_{ji}}{M_{\Xi_1}^4}\;\\
        &\left(C^{u\phi}\right)_{ij} = -\frac{2|\kappa_{\Xi_1}|^2\,(y_{u})^\ast_{ji}}{M_{\Xi_1}^4}\;.
    \end{align}
    The Lagrangian is,
    \begin{align}
        \Delta\mathcal{L} \supset &\lambda_{\Xi_1} (\Xi_1^{a\dagger}\Xi^a)(\phi^\dagger\phi) +\lambda_{\Xi_1}^{\prime} f_{abc}\,(\Xi_1^{a\dagger}\Xi^b)(\phi^\dagger\sigma^c\phi)\nonumber\\
        &(\kappa_{\Xi_1})\,\Xi_1^{a\dagger}(i\sigma^2\phi^\ast\sigma^a\phi) + \text{h.c.}\;,
    \end{align}
    where $f_{abc} = i/\sqrt{2}\varepsilon_{abc}$ and $\varepsilon_{abc}$ is the totally antisymmetric tensor.
    
    In this case, the operators are directly related to each other, we can rewrite $C^{\phi W}$ as follows, $\hat{C}^{[1]\phi W} = \hat{C}^{[1]\phi B}/3 + \frac{\lambda_{\Xi_1}}{12\sqrt{6} M_{\Xi_1}^2}$, this relation shows that we cannot easily change the sign of these two operators since they are directly related. Also strong constraints for the coupling $\kappa_{\Xi_1}$ come from the $T$-parameter. Substituting the values of the Wcs we have,
    \begin{align}
        &\Delta T = 16.033\,\frac{|\kappa_{\Xi_1}|^2}{M_{\Xi_1}^4}\;,\\
        &\Delta S = -0.008\,\frac{|\kappa_{\Xi_1}|^2}{M_{\Xi_1}^4} - 0.0019\,\frac{\lambda^\prime_{\Xi_1}}{M_{\Xi_1}^2}\;,\\
        &\delta R_{h\to\gamma\gamma} = 0.720\,\frac{|\kappa_{\Xi_1}|^2}{M_{\Xi_1}^4}\nonumber\\
        &\qquad\qquad  - \,\frac{0.0067\lambda_{\Xi_1} + 0.0033\lambda^\prime_{\Xi_1}}{M_{\Xi_1}^2}\;,\\
        &\delta R_{h\to Z\gamma} = 0.244\,\frac{|\kappa_{\Xi_1}|^2}{M_{\Xi_1}^4}\nonumber\\
        &\qquad\qquad - \,\frac{0.0015\lambda_{\Xi_1} + 0.0011\lambda^\prime_{\Xi_1}}{M_{\Xi_1}^2}\;.
    \end{align}
    Setting the T-parameter to be of the order of $T\sim 10^{-2}$, we can get a bound for the ratio $|\kappa_{\Xi_1}|^2/M_\Xi^4 \sim 0.6\times 10^{-3}$ and we rewrite the rest,
    \begin{align}
        \Delta S &\simeq -0.19\times 10^{-2}\,\frac{\lambda^\prime_{\Xi_1}}{M_{\Xi_1}^2}\;,\\
        \delta R_{h\to\gamma\gamma} &\simeq -10^{-2}\,\frac{0.67\lambda_{\Xi_1} + 0.33\lambda^\prime_{\Xi_1}}{M_{\Xi_1}^2}\;,\\
        \delta R_{h\to Z\gamma} &\simeq -10^{-2}\,\frac{0.15\lambda_{\Xi_1} + 0.11\lambda^\prime_{\Xi_1}}{M_{\Xi_1}^2}\;.
    \end{align}
    From these relation we can see that the two signal strengths cannot be separated. This model is also explored in ref \cite{Grojean:2024tcw}. The same conclusion is reached and can be seen in their Figure 6 panel (c), where the difference $\delta R_{h\to\gamma\gamma} - \delta R_{h\to Z\gamma}$ is plotted.
    \item Moving on the last two scalar fields we have a charged field labeled $\Theta_1$ in the 4-representation of $SU(2)$. The Wc expressions read,
    \begin{align}
        &\hat{C}^{[1]\phi B} = \frac{\lambda_{\Theta_1}}{16\,M_{\Theta_1}^2}\;,~ \hat{C}^{[1]\phi W} = 4\,\hat{C}^{[1]\phi B}\;,\\
        &\hat{C}^{[1]\phi WB} = - \frac{\lambda^\prime_{\Theta_1^2\phi^2}}{6\,M_{\Theta_1}^2}\;.
    \end{align}
    It is evident that we cannot under any circumstance get opposite signs for $C^{\phi B}$ and $C^{\phi W}$. Apart from the Lagrangian found in eq.(A.7) of ref. \cite{deBlas:2017xtg} we also have the additional interaction term,
    \begin{align}
        \Delta\mathcal{L} \supset \lambda^\prime_{\Theta_1^2\phi^2}\, \phi^\dagger_\alpha \Theta_{1I}^\dagger \phi_\beta \Theta_{1J}\, C^{(3)}_{\alpha I \beta J}\;,
    \end{align}
    where summation of the indices is implied. Indices range is for $\alpha,\beta = 1,2$ while for $I,J=1,2,3$. The only non-zero values of the CG tensor read
    \begin{align}
        C^{(3)}_{1112} &= C^{(3)}_{1223} = C^{(3)}_{2213} = C^{(3)}_{2221} = i\;,\\
        C^{(3)}_{1123} &= C^{(3)}_{2311} = -C^{(3)}_{1321}= -C^{(3)}_{2113} = -1\;,\\
        C^{(3)}_{1211} &= C^{(3)}_{1322} = C^{(3)}_{2122} = C^{(3)}_{2312} = -i\;.
    \end{align}
    \item The same situation stands for the other charged field labeled, $\Theta_3$, but we list here the generated operators for completeness. Part of the Lagrangian can be found on eq.(A.7) of ref. \cite{deBlas:2017xtg} and the additional part is the same as in the previous bullet, but with the substitution $\Theta_1 \to \Theta_3$ and $C^{(3)}_{\alpha I \beta J} \to C^{(4)}_{\alpha I \beta J}$. The same relations hold for $C^{(4)}_{\alpha I \beta J}$ too. The Wcs read,
    \begin{align}
        &\hat{C}^{[1]\phi B} = \frac{9\,\lambda_{\Theta_3}}{16\,M_{\Theta_3}^2}\;,~ \hat{C}^{[1]\phi W} = \frac{8}{27}\,\hat{C}^{[1]\phi B}\;,\\
        &\hat{C}^{[1]\phi WB} = -\frac{\lambda^\prime_{\Theta_3^2\phi^2}}{4\,M_{\Theta_3}^2}\;.
    \end{align}
\end{itemize}

\subsection{The case for vector-like fermions}

Barring chiral fermions, where constraints from multiple Higgs observables have ruled out this scenario \cite{Barducci:2023zml}, we can extend the fermion content of the SM by the fields shown in Table \ref{tab:fermions_ext}.

\begin{table}[ht]
    \centering
    \begin{tabular}{cccc}
        Fields & Irrep & Tree level operators & Loop level operators\\
        \hline\\
         $N$ & $(1,1)_{0}$ & $\mathcal{O}_{5}$, $\mathcal{O}_{\phi\ell}^{(1,3)}$ & $\mathcal{O}_{\phi B}$, $\mathcal{O}_{\phi W}$, $\mathcal{O}_{\phi WB}$\\
         $E$ & $(1,1)_{-1}$ & $\mathcal{O}_{e\phi}$, $\mathcal{O}_{\phi\ell}^{(1,3)}$ & $\mathcal{O}_{\phi B}$, $\mathcal{O}_{\phi W}$, $\mathcal{O}_{\phi WB}$\\
         $\Delta_{1}$ & $(1,2)_{-1/2}$ & $\mathcal{O}_{e\phi}$, $\mathcal{O}_{\phi e}$ & $\mathcal{O}_{\phi B}$, $\mathcal{O}_{\phi WB}$\\
         $\Delta_{3}$ & $(1,2)_{-3/2}$ & $\mathcal{O}_{e\phi}$, $\mathcal{O}_{\phi e}$ & $\mathcal{O}_{\phi B}$, $\mathcal{O}_{\phi WB}$\\
         $\Sigma$ & $(1,3)_{0}$ & $\mathcal{O}_{5}$, $\mathcal{O}_{e\phi}$, $\mathcal{O}_{\phi\ell}^{(1,3)}$ & $\mathcal{O}_{\phi B}$, $\mathcal{O}_{\phi W}$, $\mathcal{O}_{\phi WB}$\\
         $\Sigma_{1}$ & $(1,3)_{-1}$ & $\mathcal{O}_{e\phi}$, $\mathcal{O}_{\phi\ell}^{(1,3)}$ & $\mathcal{O}_{\phi B}$, $\mathcal{O}_{\phi W}$, $\mathcal{O}_{\phi WB}$\\
    \end{tabular}
    \caption{Tree level operators generated by new vector-like fermions.}
    \label{tab:fermions_ext}
\end{table}

The case here is clearer than the scalars because we will always need two fermions to make up interactions with the Higgs and this saturates the dimension of the corresponding operator fast.
For completeness we list the generated Wcs for each fermion listed in Table \ref{tab:fermions_ext} although none of them can accommodate to the splitting of the observables because their Wcs are proportional and the dominant ones i.e. $C^{\phi B}$ and $C^{\phi W}$ come also with the same sign:

\begin{align}
    &N:~ \hat{C}^{[1]\phi B} = \hat{C}^{[1]\phi W} = \frac{1}{2}\,\hat{C}^{[1]\phi WB} = \frac{|\lambda_{N}|^2}{24 M_{N}^{2}}\;,\\
    &E:~ \hat{C}^{[1]\phi B} = 3\,\hat{C}^{[1]\phi W} = - \frac{3}{4}\,\hat{C}^{[1]\phi WB} = \frac{|\lambda_{E}|^2}{8 M_{E}^{2}}\;,\\
    &\Delta_1:~ \hat{C}^{[1]\phi B} = -3\,\hat{C}^{[1]\phi WB}= \frac{|\lambda_{\Delta_1}|^2}{4 M_{\Delta_1}^{2}}\;,~ \hat{C}^{[1]\phi W} = 0\;,\\
    &\Delta_3:~ \hat{C}^{[1]\phi B} = 5\,\hat{C}^{[1]\phi WB}= \frac{5\,|\lambda_{\Delta_3}|^2}{12 M_{\Delta_3}^{2}}\;,~ \hat{C}^{[1]\phi W} = 0\;,\\
    &\Sigma:~ \hat{C}^{[1]\phi B} = \frac{3}{7}\,\hat{C}^{[1]\phi W}= -\frac{1}{2}\,\hat{C}^{[1]\phi WB} = \frac{|\lambda_{\Sigma}|^2}{32 M_{\Sigma}^{2}}\;,\\
    &\Sigma_1:~ \hat{C}^{[1]\phi B} = \frac{9}{7}\,\hat{C}^{[1]\phi W}= \frac{9}{8}\,\hat{C}^{[1]\phi WB} = \frac{3|\lambda_{\Sigma_1}|^2}{32 M_{\Sigma_1}^{2}}\;.
\end{align}

The Lagrangian used to obtain the Wcs for all fermions can be found in eq.(A.12) of ref. \cite{deBlas:2017xtg}, where all relevant coupling are defined. The masses of the heavy fermions are denoted by $M_{i}$, where $i=\left\{N,\,E,\,\Delta_1,\,\Delta_3,\,\Sigma,\,\Sigma_1\right\}$

Summing up, we presented all relevant Wcs, both tree and loop level, of single field extensions, of scalars and fermions, that primarily affect the signal strengths $\delta R_{h\to\gamma\gamma}$ and $\delta R_{h\to Z\gamma}$. We have found that no single field can accommodate a potential excess in one observable over the other. We must also mention that we can extend this list of fields with particles that do not have any linear coupling with the SM and leave the hypercharge as a general parameter that can be fit to find a suitable value. However, these contributions arise from the quartic interactions with the Higgs and cannot yield values that align with the single-field framework previously discussed. They become relevant when considering two-field extensions, as will be addressed in Section \ref{sec: section4}.

As a general observation, it is important to note that the value of the Wc we aim to achieve is significantly higher than what is produced in the single-field scenarios. The coefficients that scale with the coupling-to-mass ratio vary from ideally $10^{-1}$ to $10^{-3}$, these values need to be increased by one or two orders of magnitude to satisfy the requirements established in the SMEFT analysis. Addressing this challenge is complex, as the mass of the heavy field cannot be reduced significantly without compromising the convergence of the EFT, and the coupling cannot exceed $4\pi$ due to perturbativity constraints. Consequently, we are constrained to consider models involving additional fields, with the hope that their contributions may accumulate to achieve the desired effect.

\section{Two-field models}
\label{sec: section4}

In the two field case scenarios we have a plethora of combinations to work with. We can combine scalars (fermions) with scalars (fermions) and scalars with fermions and analyze each model that shows some promising characteristics on Wcs. In order to tackle the number of models that arise from these combinations we can rely on the single field results and add on top of that new scalar and/or fermion fields that could amend the situation. 

In disentangling the two-field models we will aim to categorize the interactions and their corresponding Wcs systematically, by borrowing results from functional matching techniques which directly deal with the path integral and contain a form of universality in the results, independent of the specific interaction of UV physics. In particular, when the matching at one loop is performed, and only heavy scalars run in the loop the resulting effective action is called Universal One Loop Effective Action (UOLEA) first introduced in ref. \cite{Drozd:2015rsp}. Functional matching results have also been expanded to include heavy-light particles in the loop as well as fermions \cite{Ellis:2016enq,Ellis:2017jns,Summ:2018oko,Kramer:2019fwz,Summ:2020dda,Ellis:2020ivx,Angelescu:2020yzf,Cohen:2020fcu, Dedes:2021abc}. For our purposes we will resort to results from the original UOLEA and the heavy-light UOLEA.

In order to generate the operators $C^{\phi(B,W,WB)}$, we need the functional traces to contain $G^{\prime}_{\mu\nu}\,G^{\prime}_{\mu\nu}$, where $G^{\prime}_{\mu\nu} = -i g G_{\mu\nu}$ where $g$ is the corresponding coupling of the field strength tensor $G_{\mu\nu}$, which directly relates to the gauge group representations of the field. If the fields have representations under several groups a summation over the different strength tensors is understood. There are two such traces in the heavy-only UOLEA and another two in the heavy-light UOLEA for scalar fields. The terms are shown below \cite{Dedes:2021abc},
\begin{align}
    &\tilde{f}^{9}_{i}\,\text{tr}\left\{U^{H}_{ii}\,G^{\prime}_{i,\mu\nu}\,G^{\prime}_{i,\mu\nu}\right\}\label{eq:uolea9}\;,\\
    &\tilde{f}^{13}_{ij}\,\text{tr}\left\{U^{H}_{ij}U^{H}_{ji}\,G^{\prime}_{i,\mu\nu}\,G^{\prime}_{i,\mu\nu}\right\}\;,\label{eq:uolea13}\\
    &\tilde{f}^{13A}_{i}\,\text{tr}\left\{U^{HL}_{ii^\prime}U^{LH}_{i^\prime i}\,G^{\prime}_{i,\mu\nu}\,G^{\prime}_{i,\mu\nu}\right\}\;,\label{eq:uolea13A}\\
    &\tilde{f}^{13B}_{i}\,\text{tr}\left\{U^{LH}_{i^\prime i}U^{HL}_{ii^\prime}\,G^{\prime}_{i^\prime,\mu\nu}\,G^{\prime}_{i^\prime,\mu\nu}\right\}\;,\label{eq:uolea13B}
\end{align}
where the coefficients $\tilde{f}$ read,
\begin{align}
    &\tilde{f}^{9}_{i} = -\frac{1}{12 M_i^2}\label{eq:uoleaf9}\;,\\
    &\tilde{f}^{13}_{ij} = \frac{2M_i^4 + 5 M_i^2 M_j^2 - M_j^4}{12 M_i^2 (\Delta_{ij}^2)^3} - \frac{M_i^2 M_j^2}{2(\Delta_{ij}^2)^4}\,\log\left(\frac{M_i^2}{M_j^2}\right)\;,\\
    &\tilde{f}^{13A}_{i} = \frac{1}{6 M_i^4}\;,\\
    &\tilde{f}^{13B}_{i} = -\frac{1}{4 M_i^4}\;,
\end{align}

where $M_{i}$ denotes the mass of the heavy scalar field. The coefficients are defined as $\tilde{f} = 16\pi^2\,f$ and are obtained through integration of Feynman integrals over momentum. They are labelled as universal coefficients because they factor out the mass dependence of the UV physics. The unprimed indices of the $U$-matrices denote the set of heavy fields, while the primed indices denote the set of light fields. We have also labeled the mass difference of the heavy fields as $\Delta_{ij}^2 = M_i^2 -M_j^2$ for brevity. The $U$-matrices are nothing more than differentiation of the Lagrangian with respect to the fields that the subscript indices denote, hence they contain both fields and couplings which in the end need to be traced throughout all available spaces of the fields (i.e. flavor, SU(2), Lorentz etc.).

We can now categorize the interactions needed to produce our Wcs of interest through the terms in eqs. (\ref{eq:uolea9}-\ref{eq:uolea13B}). For example, in the first term (\ref{eq:uolea9}) the mass dimension of $U$ is $[U^{H}_{ii}] = 2$, which must contain two Higgs fields to produce the desired operators, and it has been differentiated two times with respect to the heavy field, thus we can schematically write down the Lagrangian term as, $\Delta \mathcal{L} \sim X_{i}X_{i} H^\dagger H$. Following the same line of thought for the other two terms we can then write the correspondence,
\begin{align}
    &\tilde{f}^{9}_{i}\,\text{tr}\left\{U^{H}_{ii}\,G^{\prime}_{i,\mu\nu}\,G^{\prime}_{i,\mu\nu}\right\} \longrightarrow \Delta \mathcal{L} \sim X_{i}X_{i} H^\dagger H\;,\label{eq:lag9}\\
    &\tilde{f}^{13}_{ij}\,\text{tr}\left\{U^{H}_{ij}U^{H}_{ji}\,G^{\prime}_{i,\mu\nu}\,G^{\prime}_{i,\mu\nu}\right\} \longrightarrow \Delta\mathcal{L} \sim X_i X_j H + \text{h.c.}\;,\label{eq:lag13}\\
    &\begin{rcases}
        \tilde{f}^{13A}_{i}\,\text{tr}\left\{U^{HL}_{ii^\prime}U^{LH}_{i^\prime i}\,G^{\prime}_{i,\mu\nu}\,G^{\prime}_{i,\mu\nu}\right\}\\
        \tilde{f}^{13B}_{i}\,\text{tr}\left\{U^{LH}_{i^\prime i}U^{HL}_{ii^\prime}\,G^{\prime}_{i^\prime,\mu\nu}\,G^{\prime}_{i^\prime,\mu\nu}\right\}
    \end{rcases}
    \longrightarrow \Delta\mathcal{L} \sim X_i X_{i^\prime} H + \text{h.c.}\;.\label{eq:lag13ab}
\end{align}

We note that for (\ref{eq:lag13ab}), the only light scalar in the SM is the Higgs thus $X_{i^\prime} = H$ and the interaction becomes $\Delta \mathcal{L} \sim X_i H^\dagger H$ and the only suitable fields have hypercharge $Y_i = 0$ and correspond to the neutral singlet $S$ and the triplet $\Xi$, which were previously discussed. We can then pair up these two fields with any other one to give us the desired operators. Next, eq.(\ref{eq:lag9}) conserves the hypercharge universally, that is this term is going to be present even if the value of the hypercharge is arbitrary. Lastly, eq.(\ref{eq:lag13}) is a linear coupling of the Higgs and the heavy fields so the hypercharge of the corresponding new fields have to obey the following rule, if we suppose that we leave $Y_i$ free, the hyperchage of $X_j$ is going to take the value $Y_j = Y_i + Y_H$ with $Y_H=1/2$. In Table \ref{tab:two_scalars_models} we present the two-field models that could generate the operators $C^{\phi(B,W,WB)}$. We split the table depending on the field content, either we have two scalars labelled as $(SS)$, a scalar and a fermion labeled as $(SF)$ or two fermions labeled as $(FF)$.

\begin{table*}
    \centering
    \begin{tabular}{c|ccc}
        Model \# & Field 1 & Field 2 & Tree Level \\
        \hline\\
        $SS_{10 1Y_2}$ & $S_a \to (1,1,0)$ & $S_b \to (1,1,Y_2)$ & $\mathcal{O}^{\phi\Box}$\\
        $SS_{10 2Y_2}$ & $S_a \to (1,1,0)$ & $S_b \to (1,2,Y_2)$ & $\mathcal{O}^{\phi\Box}$\\
        $SS_{10 3Y_2}$ & $S_a \to (1,1,0)$ & $S_b \to (1,3,Y_2)$ & $\mathcal{O}^{\phi\Box}$\\
        $SS_{30 2Y_2}$ & $S_a \to (1,3,0)$ & $S_b \to (1,2,Y_2)$ & $\mathcal{O}^{\phi\Box}$, $\mathcal{O}^{\phi D}$, $\mathcal{O}^{(e,u,d)\phi}$\\
        $SS_{30 3Y_2}$ & $S_a \to (1,3,0)$ & $S_b \to (1,3,Y_2)$ & $\mathcal{O}^{\phi\Box}$, $\mathcal{O}^{\phi D}$, $\mathcal{O}^{(e,u,d)\phi}$\\
        \hline\\
        $SS_{1Y_1 2Y_2}$ & $S_a \to (1,1,Y_1)$ & $S_b \to (1,2,Y_1+1/2)$ & -\\
        $SS_{2Y_1 1Y_2}$ & $S_a \to (1,2,Y_1)$ & $S_b \to (1,1,Y_1+1/2)$ & -\\
        $SS_{2Y_1 3Y_2}$ & $S_a \to (1,2,Y_1)$ & $S_b \to (1,3,Y_1+1/2)$ & -\\
        $SS_{3Y_1 2Y_2}$ & $S_a \to (1,3,Y_1)$ & $S_b \to (1,2,Y_1+1/2)$ & -\\
        \hline\\
        $SF_{10 1Y_2}$ & $S_a \to (1,1,0)$ & $F_b \to (1,1,Y_2)$ & $\mathcal{O}^{\phi\Box}$\\
        $SF_{10 2Y_2}$ & $S_a \to (1,1,0)$ & $F_b \to (1,2,Y_2)$ & $\mathcal{O}^{\phi\Box}$\\
        $SF_{10 3Y_2}$ & $S_a \to (1,1,0)$ & $F_b \to (1,3,Y_2)$ & $\mathcal{O}^{\phi\Box}$\\
        $SF_{30 2Y_2}$ & $S_a \to (1,3,0)$ & $F_b \to (1,2,Y_2)$ & $\mathcal{O}^{\phi\Box}$, $\mathcal{O}^{\phi D}$, $\mathcal{O}^{(e,u,d)\phi}$\\
        $SF_{30 3Y_2}$ & $S_a \to (1,3,0)$ & $F_b \to (1,3,Y_2)$ & $\mathcal{O}^{\phi\Box}$, $\mathcal{O}^{\phi D}$, $\mathcal{O}^{(e,u,d)\phi}$\\
        \hline\\
        $FF_{1Y_1 2Y_2}$ & $F_a \to (1,1,Y_1)$ & $F_b \to (1,2,Y_1+1/2)$ & -\\
        $FF_{2Y_1 1Y_2}$ & $F_a \to (1,2,Y_1)$ & $F_b \to (1,1,Y_1+1/2)$ & -\\
        $FF_{2Y_1 3Y_2}$ & $F_a \to (1,2,Y_1)$ & $F_b \to (1,3,Y_1+1/2)$ & -\\
        $FF_{3Y_1 2Y_2}$ & $F_a \to (1,3,Y_1)$ & $F_b \to (1,2,Y_1+1/2)$ & -\\
    \end{tabular}
    \caption{Two field models, all singlets under SU(3), that can generate $C^{\phi(B,W,WB)}$. We have established the following naming convention for the models. Capital letters denote field content as described in the text. The first two subscript indices denote gauge quantum numbers under $SU(2) \times U(1)$ for the first field shown in column 2 (Field 1), while the last two denote the gauge quantum numbers under $SU(2) \times U(1)$ for the field in column 3 (Field 2). In the subscripts we refrain from denoting $SU(3)$ since all field under consideration are singlets under this gauge group. In columns 2 and 3, the full representation of the gauge group, under $SU(3)\times SU(2) \times U(1)$ is shown for the fields considered in the model of column 1. In total there are 18 candidate models.} 
    \label{tab:two_scalars_models}
\end{table*}

Our strategy to determine the most we can get in the decay $h\to Z\gamma$ will be to construct a $\chi^2$ with all observables mentioned in Table \ref{tab:numerics}, excluding $h \to Z\gamma$ and leaving it as prediction of our model. The main reason for substituting the decay $h\to Z\gamma$ from a constraint to a prediction is that we want more freedom in the parameter space of each model, and leaving this specific decay as a models' prediction is more suitable for our purpose. Meanwhile, it yields slightly better results than having it as a constraint. Seeking a sign difference in the fashion of Section \ref{sec: section3} is not an option for this procedure in certain models that we are going to investigate. The reason behind this, is that Wcs expressions in specific cases become much more complex and an automated procedure is deemed more useful. Also, Wcs serve as a medium for the couplings of each UV model. Since we are now considering UV physics with definite couplings it is more fit to discuss couplings directly related to physical observables instead of Wcs. However, one can seek the values of Wcs presented in Section \ref{sec: section2} given for each model in the auxiliary csv file.

We set bounds to the relevant couplings and masses and perform a constrained minimization of the $\chi^2$ for each model. In the minimization we bound the couplings of each model to a range of $\lambda \sim [-2,2]$ so as to account for perturbativity and for the masses we set the lowest values to be $M_{S(F)} \geq 0.7\,\text{TeV}$ to avoid issues with the EFT validity, since the lowest scale is the Higgs vev. In the models where the hypercharge relation obeys $Y_j = Y_i +1/2$, the minimization with respect to the masses is more involved since there is also the degenerate mass limit that needs to be accounted for. In the non-degenerate case we restrict the mass difference to be greater than $0.1~\text{TeV}$ keeping of course the lowest bound of $0.7~\text{TeV}$ for both masses. We have also explored the degenerate mass limit, where in some cases we saw a slight improvement while in other cases we saw the gap between the two observables widen. Finally, we substitute the best fit values into $\delta R_{h\to Z\gamma}$ to get the models' prediction. The values of the corresponding Wcs at both tree and loop-level are matched with MatchMakerEFT \cite{Carmona:2021xtq} and, for exclusively loop generated operators, the expressions from MatchMakerEFT are also cross-checked with the package SOLD \cite{Guedes:2023azv}.

\begin{figure*}
    \includegraphics[scale=0.44]{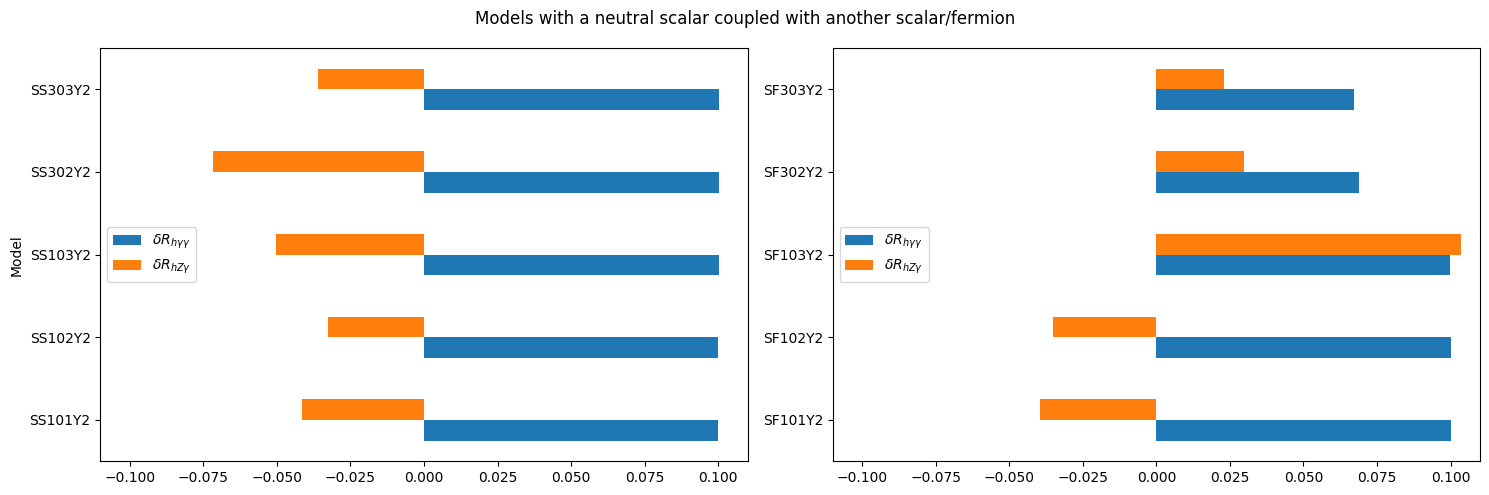}
    \vspace{.2cm}
    \includegraphics[scale=0.44]{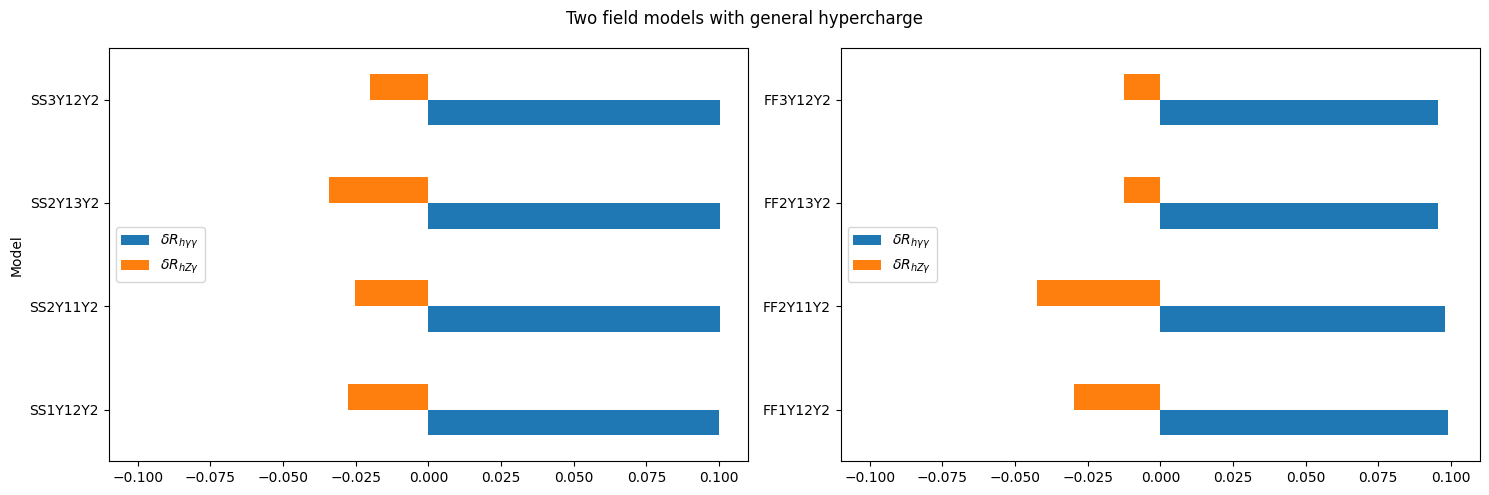}
    \caption{The orange color represents the prediction of $h\to Z\gamma$ for each model while blue represents the values of $h\to\gamma\gamma$ obtained from the minimization of the couplings and masses (in the non-degenerate limit).}
    \label{fig:results}
\end{figure*}

In Figure \ref{fig:results} we present the results of the minimization procedure along with the prediction of $h\to Z\gamma$ of each respective model and the value of $h\to\gamma\gamma$ obtained by this minimization. We observe that only one model predicts the value of $\delta R_{h\to Z\gamma}$ to be greater than that of $\delta R_{h\to\gamma\gamma}$. The only model capable to boost the one over the other observable, albeit with a negligible difference is $SF_{103Y_2}$, containing a neutral scalar singlet and a fermion triplet. The masses of the particles provided by the minimization subject to the constrains mentioned previously are for the scalar $M_S = 1.62\text{ TeV}$ and for the fermion $M_F = 0.7\text{ TeV}$, with a hypercharge $Y_2 = 0$. The exact values of the prediction is $\delta R_{h\to\gamma\gamma} = 0.099$ and $\delta R_{h\to Z\gamma} = 0.104$. Their difference is of the order $\sim 0.5\%$ with respect to the SM signal strength $\mu = 1$ and with the assumptions we have made on the models it can never reach the observed difference of the two signal strengths.

\section{Conclusions}
\label{sec: section5}

Following the first evidence of the Higgs boson decaying into a photon and a Z-boson by the ATLAS and CMS collaborations, we addressed the reported excess of $\sim 2\sigma$ \cite{ATLAS:2023yqk}. We conducted a model-independent analysis in the SMEFT, incorporating various observables in both the decay and production channels of the Higgs boson. By performing a $\chi^2$-minimization, we found the best-fit values for each Wilson coefficient. This procedure reveals the most general relations among the Wcs to account for any discrepancy. Our main finding is that the Wilson coefficients $C^{\phi B}$ and $C^{\phi W}$ need to have opposite signs and have comparable magnitudes, while $C^{\phi WB}$ needs to be small since it's heavily constrained by the $S$-parameter.

Using a model-independent approach, we identified the necessary characteristics that UV models must possess to account for the excess in $h\to Z \gamma$. We considered all scalar and fermion single-field extensions of the SM that respect the SM gauge group but found that none could accommodate the data. Subsequently, we examined two-field models combining scalars and/or fermions. The candidate models were categorized based on their content and their universal loop-integral coefficient using the UOLEA.

We matched all UV models using automated packages and constructed a new $\chi^2$ function, which we minimized with respect to model parameters. As shown in Figure \ref{fig:results}, out of the 18 model families, only one specific model, which includes a neutral scalar singlet and a neutral fermion triplet, can boost $h \to Z\gamma$ while preserving the $h\to\gamma\gamma$ decay. However, this model still cannot fully accommodate the observed discrepancy in the data.

In summary, this study is valuable for providing the values and characteristics of Wcs capable of generating an enhancement in the decay channel $h\to Z\gamma$ over $h\to\gamma\gamma$. Moreover, the exploration of various two-field model families, surpassing the limitations of single-field models, provides insightful perspectives into potential UV physics behind this mild excess in $h\to Z\gamma$. If future data confirm this excess, our attention can pivot towards the investigation of alternative UV physics as the source of the discrepancy.

\vspace{.5cm}
\section*{Acknowledgements}
I would like to thank A. Dedes for his helpful suggestions on the manuscript and for his numerous discussions during the preparation of this work.

\appendix
\newpage
\section{Semi-numerical expressions of signal strengths}
\label{app: A}

Here we provide the semi-numerical expressions of the signal strength corrections to the SM, in the decays and production channels of the Higgs boson, that we have used in this study. Apart from $\delta R_{h\to Z\gamma}$ and $\delta R_{h\to\gamma\gamma}$, which their one loop expressions are considered, found in \cite{Dedes:2018seb,Dedes:2019bew}, all other formulas are taken to leading order from \cite{DasBakshi:2024krs} and are computed in $\rm{TeV}^{-2}$ units. All expressions are given in the input scheme $\{G_{F}, M_{W}, M_{Z}\}$, the numerical values used throughout this work are,

\begin{align}
    G_{F} &= 1.1663787 \times 10^{-5}~\text{GeV}^{-2}\;,\\
    M_{W} &= 80.385~\text{GeV}\;,\\
    M_{Z} &= 91.1876~\text{GeV}\;,\\
    M_{H} &= 125.25~\text{GeV}\;,\\
    M_{t} &= 172.57~\text{GeV}\;.
\end{align}

\subsection{Decay channels}

\begin{align}
    &\delta R_{h\to b \bar{b}} = -5.050\, C^{d\phi}_{33} + 0.121\, (C^{\phi\Box} - \frac{1}{4} C^{\phi D}) + 0.0606\, (C^{\ell\ell}_{1221} - C^{\phi\ell(3)}_{11} -C^{\phi\ell(3)}_{22})\;,\\
    &\delta R_{h\to WW^\ast} = -0.0895\, C^{\phi W} + 0.121\, (C^{\phi\Box} - \frac{1}{4} C^{\phi D}) + 0.0606\, (C^{\ell\ell}_{1221} - C^{\phi\ell(3)}_{11} -C^{\phi\ell(3)}_{22})\;,\\
    &\delta R_{h\to \tau \bar{\tau}} = -11.88\, C^{e\phi}_{33} + 0.121\, (C^{\phi\Box} - \frac{1}{4} C^{\phi D}) + 0.0606\, (C^{\ell\ell}_{1221} - C^{\phi\ell(3)}_{11} -C^{\phi\ell(3)}_{22})\;,\\
    &\delta R_{h\to ZZ^\ast} = 0.296\,(C^{\phi WB} - C^{\phi W}) - 0.197\, C^{\phi B} + 0.119\, C^{\phi\Box} + 0.005 C^{\phi D} + 0.181\,C^{\ell\ell}_{1221}\nonumber\\
    &- 0.117\,(C^{\phi\ell(3)}_{11} + C^{\phi\ell(3)}_{22})\;,\\
    & \delta R_{h\to \mu\mu} = -199.79\, C^{e\phi}_{22} + 0.121\, (C^{\phi\Box} - \frac{1}{4} C^{\phi D})\;,\\
    \nonumber\\
    &\delta R_{h \to Z \gamma} \simeq 0.18 \left(C^{\ell \ell}_{1221} -C^{\phi \ell (3)}_{11} -C^{\phi \ell (3)}_{22} \right) + 0.12 \left( C^{\phi\Box} -C^{\phi D} \right) \nonumber\\
    &- 0.01 \left( C^{d\phi}_{33} - C^{u\phi}_{33} \right) + 0.02 \left( C^{\phi u}_{33} +C^{\phi q(1)}_{33} -C^{\phi q(3)}_{33} \right)\nonumber\\
    &+\left[14.99 - 0.35 \log\frac{\mu^2}{M_{W}^{2}} \right] C^{\phi B} - \left[14.88 - 0.15 \log\frac{\mu^2}{M_{W}^{2}} \right] C^{\phi W} +\left[9.44 - 0.26 \log\frac{\mu^2}{M_{W}^{2}} \right] C^{\phi WB}\nonumber\\
    &+ \left[0.10 - 0.20 \log\frac{\mu^2}{M_{W}^{2}} \right] C^{W} -\left[0.11 - 0.04 \log\frac{\mu^2}{M_{W}^{2}} \right] C^{uB}_{33} + \left[0.71 - 0.28 \log\frac{\mu^2}{M_{W}^{2}} \right] C^{uW}_{33}\nonumber\\
    &-0.01\, C^{uW}_{22} - 0.01\,C^{dW}_{33} + \ldots\;,\\
    \nonumber\\
    &\delta R_{h \to \gamma \gamma} \simeq 0.18 \left(C^{\ell \ell}_{1221} -C^{\phi \ell (3)}_{11} -C^{\phi \ell (3)}_{22} \right) + 0.12 \left( C^{\phi\Box} - 2C^{\phi D} \right) \nonumber\\
    &- 0.01 \left( C^{e\phi}_{22} + 4C^{e\phi}_{33}  +5C^{u\phi}_{22} +2C^{d\phi}_{33} -3C^{u\phi}_{33} \right)\nonumber\\
    &-\left[48.04 - 1.07 \log\frac{\mu^2}{M_{W}^{2}} \right] C^{\phi B} - \left[14.29 - 0.12 \log\frac{\mu^2}{M_{W}^{2}} \right] C^{\phi W} +\left[26.17 - 0.52 \log\frac{\mu^2}{M_{W}^{2}} \right] C^{\phi WB}\nonumber\\
    &+ \left[0.16 - 0.22 \log\frac{\mu^2}{M_{W}^{2}} \right] C^{W} +\left[2.11 - 0.84 \log\frac{\mu^2}{M_{W}^{2}} \right] C^{uB}_{33} + \left[1.13 - 0.45 \log\frac{\mu^2}{M_{W}^{2}} \right] C^{uW}_{33}\nonumber\\ &-\left[0.03 - 0.01 \log\frac{\mu^2}{M_{W}^{2}} \right] C^{uB}_{22}
    - \left[0.01 - 0.00 \log\frac{\mu^2}{M_{W}^{2}} \right] C^{uW}_{22}
    +\left[0.03 - 0.01 \log\frac{\mu^2}{M_{W}^{2}} \right] C^{dB}_{33}\nonumber\\
    &-\left[0.02 - 0.01 \log\frac{\mu^2}{M_{W}^{2}} \right] C^{dW}_{33} +\left[0.02 - 0.00 \log\frac{\mu^2}{M_{W}^{2}} \right] C^{eB}_{33} - \left[0.01 - 0.00 \log\frac{\mu^2}{M_{W}^{2}} \right] C^{eW}_{33} + \ldots\;.
\end{align}

\subsection{Production channels}

\begin{align}
    &\delta R_{\rm{ggF}} = 0.249\, C^{d\phi}_{33} + 0.121\, C^{\phi\Box} - 0.303\, C^{\phi D} - 0.129\, C^{u\phi}_{33} - 0.0606 (C^{\phi\ell(3)}_{11} + C^{\phi\ell(3)}_{22} - C^{\ell\ell}_{1221})\;,\\
    &\delta R_{\rm{VBF}} = -0.423\, C^{\phi q(3)}_{11} - 0.347\, C^{\phi q(1)}_{11} +0.1005 C^{\phi\Box} + 0.0826\, C^{\ell\ell}_{1221} - 0.0670\, C^{\phi W}\nonumber\\
    &- 0.0150\, C^{\phi D} + 0.0126 C^{\phi WB} - 0.0107 C^{\phi B}\;,\\
    &\delta R_{Wh} = 1.950\, C^{\phi q(3)}_{11} + 0.887\, C^{\phi W} + 0.127\, C^{\phi\Box} +0.0606\, C^{\ell\ell}_{1221} - 0.0303\, C^{\phi D}\;,\\
    &\delta R_{Zh} = 1.716\, C^{\phi q(3)}_{11} +0.721\, C^{\phi W} + 0.426\, C^{\phi u}_{11} - 0.173\, C^{\phi q (1)}_{11} - 0.142\, C^{\phi d}_{11} + 0.121\, C^{\phi\Box}\nonumber\\
    &+ 0.0865\, C^{\phi B} + 0.0375\, C^{\phi D} + 0.314\, C^{\phi WB} + 0.06045\, C^{\ell\ell}_{1221}\;,\\
    &\delta R_{t\bar{t}h} = 0.121\, C^{\phi\Box} - 0.122\, C^{u\phi}_{33} - 0.0606(C^{\phi\ell (3)}_{11} + C^{\phi\ell (3)}_{22} - C^{\ell\ell}_{1221}) - 0.0303\, C^{\phi D}\;.
\end{align}

\section{Model files and Wilson coefficients}
\label{app: B}

As a complementary material we provide a csv file containing the two field models and several values of Wcs, observables and coupling values. More specifically the columns of the csv are the following 

\begin{align}
    &\left\{Model\right.\\
    &\left.L1,\, L1bar,\, L2,\, L3,\, L4,\, L5,\, L6,\, L7,\, L8,\, L9,\, L10,\, L1L,\, L1Lbar,\, L1R,\, L1Rbar,\, L5L,\, L5R,\right.\\
    &\left. MSa,\, MSb,\, MFa,\, MFb,\, Y2,\right.\\
    &\left. CphiBox,\, CphiD,\, Cuphi33,\, Cdphi33,\, CphiB,\, CphiW,\, CphiWB,\right.\\
    &\left. dRhgamgam,\, dSparam,\, dTparam,\, dRhbb,\, dRhWW,\, dRhtautau,\, dRhZZ,\,dRggF,\right.\\
    &\left. dRVBF,\, dRWh,\, dRZh,\, dRtth,\, dRhmumu,\, dRhZgam,\, chi2min\right\}
\end{align}

In the first column is the model name as we discussed in Table \ref{tab:two_scalars_models}. The next columns whose labels begin with $L$ are the various couplings stemming from the Lagrangian of the model files and their corresponding values upon minimization of the $\chi^2$. Following are the relevant masses and the hypercharge for each model. Next, are the relevant Wcs, production and decay rates in this study and their values using the best-fit coupling from the previous columns. Finally, $dRhZgam$ contains the predictions to $\delta R_{h\to Z\gamma}$ for each respective model and for clarity we also mention in the last column the value of $\chi^2_{\text{min}}$. Some models may not depend on all the quantities listed above. For example, the two scalar models do not depended on any fermion masses and vice versa. These values are treated as $NaN$ values in the csv. One can load the csv in a Dataframe using python and the corresponding package pandas with the following lines of code,
    \begin{verbatim}
        import pandas as pd
    
        df = pd.read_csv(
            "your-path-to-csv/all_models.csv")
    \end{verbatim}

Additionally to the csv, we include all models files generated by SOLD, through its interface with the package MatachMakerEFT for the two-field model cases presented in this work as a zip file. In the model files one can find the connection between the couplings in the csv and the actual interactions in the Lagrangian. 

\bibliography{refs.bib}
\bibliographystyle{utphys.bst}

\end{document}